\documentclass[prl,twocolumn,Letters, superscriptaddress]{revtex4}

\usepackage{graphicx}
\newcommand{\beq}{\begin{equation}}
\newcommand{\eeq}{\end{equation}}
\newcommand{\beqa}{\begin{eqnarray}}
\newcommand{\eeqa}{\end{eqnarray}}

\newcommand{\ket}[1]{\mbox{$ | #1 \rangle $}}

\def\opone{\leavevmode\hbox{\small1\normalsize\kern-.33em1}}

\begin{document}

\title{Why Bohmian Mechanics? one and two-time position measurements, Bell inequalities, philosophy and physics}

\author{Nicolas Gisin \\
\it \small   Group of Applied Physics, University of Geneva, 1211 Geneva 4,    Switzerland}

\date{\small \today}

\begin{abstract}
In Bohmian mechanics particles follow continuous trajectories, hence 2-time position correlations are well defined. Nevertheless, Bohmian mechanics predicts the violation of Bell inequalities. Motivated by this fact we investigate position measurements in Bohmian mechanics by coupling the particles to macroscopic pointers. This explains the violation of Bell inequalities despite 2-time position correlations. We relate this fact to so-called surrealistic trajectories that, in our model, correspond to slowly moving pointers. Next, we emphasize that the nice feature of Bohmian mechanics, which doesn't distinguish microscopic and macroscopic systems, implies that the quantum weirdness of quantum physics also shows up at the macro-scale. Finally, we discuss the fact that Bohmian mechanics is attractive to philosophers, but not so much to physicists and argue that the Bohmian community is responsible for the latter.
\end{abstract}

\maketitle

\section{Introduction}
%=======================
Bohmian mechanics differs deeply from standard quantum mechanics. In particular, in Bohmian mechanics particles, here called Bohmian particles, follow continuous trajectories; hence in Bohmian mechanics there is a natural concept of time-correlation for particles' positions. This led M. Correggi and G. Morchio \cite{CorreggiMorchio02} and more recently Kiukas and Werner \cite{KiukasWerner09} to conclude that Bohmian mechanics ``can't violate any Bell inequality'', hence is disproved by experiments. However, the Bohmian community maintains its claim that Bohmian mechanics makes the same predictions as standard quantum mechanics (at least as long as only position measurements are considered, arguing that, at the end of the day, all measurements result in position measurement, e.g. pointer's positions).

Here we clarify this debate. First, we recall why two-time position correlation is at a tension with Bell inequality violation. Next, we show that this is actually not at odd with standard quantum mechanics because of some subtleties. For this purpose we do not go for full generality, but illustrate our point on an explicit and rather simple example based on a two-particle interferometers, partly already experimentally demonstrated and certainly entirely experimentally feasible (with photons, but also feasible at the cost of additional technical complications with massive particles). The subtleties are illustrates by explicitly coupling the particles to macroscopic systems, called pointers, that measure the particles' positions. Finally, we raise questions about Bohmian positions, about macroscopic systems and about the large difference in appreciation of Bohmian mechanics by the philosophers and physicists communities.

\section{Bohmian positions}
%==========================
Bohmian particles have, at all times, well defined positions in our 3-dimensional space. However, for the purpose of my analysis I need only to specify in which mode the Bohmian particle is. For example, if a particle in mode 1 encounters a beam-splitter with output modes 1 and 2, then the Bohmian particle exits the beam-splitter either in mode 1 or in mode 2, see Fig. 1. 

\begin{figure}
\centering
\includegraphics[width=9cm]{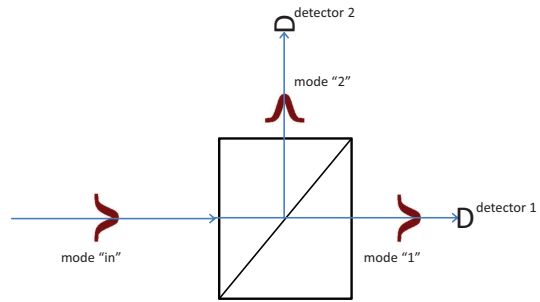}
\caption{A Bohmian particle and its pilot wave arrive on a Beam-Splitter (BS) from the left in mode ``in''. The pilot wave emerges both in modes 1 and 2, as the quantum state in standard quantum theory. However, the Bohmian particle emerges either in mode 1 or in mode 2, depending on its precise initial position. As Bohmian trajectories can't cross each other, if the initial position is in the lower half of mode ``in'', then the Bohmian particle exists the BS in mode 1, else in mode 2.}\label{fig1}
\end{figure}

Part of the attraction of Bohmian mechanics lies then in the assumption that 
\begin{itemize}
	\item 
	\underline{Assumption {\bf H} :} \\{\it Position measurements merely reveal in which (spatially separated and non-overlapping) mode the Bohmian particle actually is}. \label{assump}
\end{itemize}

Accordingly, if the modes 1 and 2 after the beam-splitter are connected to two single-particle detectors, then if the Bohmian particle is in mode 1, the corresponding detector click, and similarly for the case of mode 2, see Fig. 1.

\section{Two-time position correlation in a Bell test}
%=====================================================
Let's consider a 2-particle experiment with 4 modes, labelled 1,2,3 and 4, as illustrated in Fig. 2. The source produces the quantum state:
\beq\label{psi_in}
\psi_0=\big(\ket{1001}+\ket{0110}\big)/\sqrt{2}
\eeq
where, e.g., $\ket{1001}$means that there is one particle in mode 1 and one in mode 4, with modes 2 and 3 empty. This is an entangled state that can be used in a Bell inequality test. For this, Alice (who controls modes 1 and 2) and Bob (who controls modes 3 and 4) apply phases $x$ and $y$ to modes 1 and 4, respectively, and combine their modes on a beam-splitter, see Fig. 2. Taking into account that a reflection on a BS induces a phase $e^{i\pi/2}=i$, the quantum state after the two beam-splitters reads:
\beqa
\frac{e^{i(x+y)}}{2^{3/2}}\big(\ket{1001}+i\ket{0101}+i\ket{1010}-\ket{0110}\big) \nonumber\\
+~\frac{1}{2^{3/2}}\big(\ket{0110}+i\ket{0101}+i\ket{1010}-\ket{1001}\big)
\eeqa
If the modes 1,2,3 and 4 after the beam-splitter encounter 4 single-particle detectors, also labelled 1,2,3 and 4, then the probabilities for coincidence detection read:
\beqa\label{Qcorrel1}
P_{14}=P_{23}=\frac{1}{8}|e^{i(x+y)}-1|^2=\frac{1-\cos(x+y)}{4}\\
P_{13}=P_{24}=\frac{1}{8}|e^{i(x+y)}+1|^2=\frac{1+\cos(x+y)}{4} \label{Qcorrel2}
\eeqa
from which a maximal violation of the CHSH-Bell inequality can be obtained with appropriate choices of the phase inputs.

\begin{figure}
\centering
\includegraphics[width=7cm]{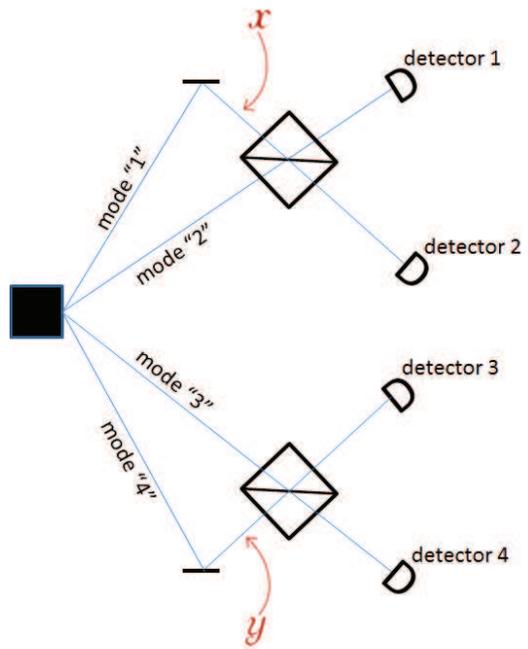}
\caption{Two Bohmian particles spread over 4 modes. The quantum state is entangled, see eq. (\ref{psi_in}), hence the two particle are either in modes 1 and 4, or in modes 2 and 3. Alice applies a phase $x$ on mode 1 and Bob a phase $y$ on mode 4. Accordingly, after the two beam-splitters the correlations between the detectors allow Alice and Bob to violate Bell inequality. The convention on mode numbering is that modes don't cross, i.e. the $n$th mode before the beam-splitter goes to detector $n$.}\label{fig2}
\end{figure}

In Bohmian mechanics this experiment is easily described. Denote the two particles' positions $r_A$ and $r_B$. In the initial state (\ref{psi_in}) the particles are either in modes 1 and 4, a situation we denote $r_A\in\,''1''$ and $r_B\in\,''4''$, or in modes 2 and 3, i.e. $r_A\in\,''2''$ and $r_B\in\,''3''$. According to Bohmian mechanics the particles have more precise positions, but for our argument this suffices.

Now, according to Bohmian mechanics and assumption {\bf H}, one doesn't need to actually measure the positions of the particles, it suffices to know that each is in one specific mode. Hence, one can undo Alice's measurement as illustrated in Fig. 3. After the phase shift $-x$ the quantum state is precisely back to the initial state $\psi_0$, see (\ref{psi_in}). Alice can thus perform a second measurement with a freshly chosen phase $x'$ and a third beam-splitter, see fig. 3. Moreover, as Bohmian trajectories can't cross each others, if $r_A$ was in mode 1 before the first BS, then $r_A$ is also in mode 1 before the last BS.

\begin{figure}
\centering
\includegraphics[width=9cm]{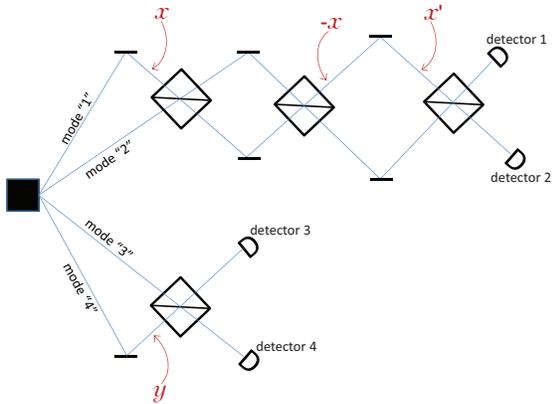}
\caption{Alice's first ``measurement'', with phase $x$, can be undone because in Bohmian mechanics there is no collapse of the wavefunction. Hence, after having applied the phase $-x$ after her second beam-splitter, Alice can perform a second ``measurement'' with phase $x'$. The mode number convention implies, e.g., that mode 1 is always the upper mode, i.e. the mode on which all phases $x$, $-x$ and $x'$, are applied.}\label{fig3}
\end{figure}

There is no doubt that according to Bohmian mechanics there is a well-defined joint probability distribution for Alice's particle at two times and Bob's particle: $P(r_A,r'_A,r_B|x,x',y)$, where $r_A$ denotes Alice's particle after the first beam-splitter and $r'_A$ after the third beam-splitter of Fig. 3. But here comes the puzzle. According to Assumption {\bf H}, if $r_A\in\,''1''$, then any position measurement performed by Alice in-between the first and second beam-splitter would necessarily result in $a=1$. Similarly $r_A\in\,''2''$ implies $a=2$. And so on, Alice's position measurement after the third beam-splitter is determined by $r'_A$ and Bob's measurement determined by $r_B$. Hence, it seems that one obtains a joint probability distribution for both of Alice's measurements results and for Bob's: $P(a,a',b|x,x',y)$. But such a joint probability distribution implies that Alice doesn't have to make any choice (she merely makes both choices, one after the other), and in such a situation there can't be any Bell inequality violation. Hence, as claimed in \cite{KiukasWerner09}, it seems that the existence of 2-time position correlations in Bohmian mechanics prevents the possibility of a CHSH-Bell inequality violation, in contradiction with quantum theory predictions and experimental demonstrations \cite{RMP-NL-Brunner14}.

Let's have a closer look at the probability distribution that lies at the bottom of our puzzle: $P(r_A,r'_A,r_B|x,x',y)$. More precisely, it suffices to consider in which modes the Bohmian particles are. That is, it suffices to consider the following joint probability distribution:
\beq
P(r_A\in``a``,r'_A\in``a'``,r_B\in``b``|x,x',y)
\eeq
where $a,a'=1,2$ and $b=3,4$.
This can be computed explicitly:
\beqa
P(r_A\in``a``,r'_A\in``a'``,r_B\in``b``|x,x',y)=\nonumber\\
\frac{1+(-1)^{a+b}\cos(x+y)}{4}\cdot\frac{1+(-1)^{a'+b}\cos(x+y)}{2} \label{jointProb}
\eeqa

Note that if one sums over $a'$, i.e. traces out Alice's second measurement, then one recovers the quantum prediction (\ref{Qcorrel1}) and (\ref{Qcorrel2}):
\beqa\label{correlab}
P(r_A\in``a``,r_B\in``b``|x,y)=\nonumber\\
\sum_{a'}P(r_A\in``a``,r'_A\in``a'``,r_B\in``b``|x,x',y)=\nonumber\\
\frac{1+(-1)^{a+b}\cos(x+y)}{4}
\eeqa
Important is to notice that $P(r_A\in``a``,r_B\in``b``|x,y)$ does not depend on Alice's second measurement setting $x'$, as one should expect.
Similarly, if one traces out Alice's first measurement:
\beqa\label{correla'b}
P(r_A\in``a'``,r_B\in``b``|x',y)=\nonumber\\
\frac{1+(-1)^{a'+b}\cos(x'+y)}{4}
\eeqa
one recovers  (\ref{Qcorrel1}) and (\ref{Qcorrel2}). Again, the probability (\ref{correla'b}) doesn't depend on Alice's first measurement setting.

So far so good, but now comes the catch. If one traces out Bob's measurement one obtains a probability distribution for Alice's particle's position that depends on Bob's setting $y$:
\beqa\label{PosCorrel}
P(r_A\in``a``,r_A\in``a'``|x,x',y)=\nonumber\\
\sum_bP(r_A\in``a``,r'_A\in``a'``,r_B\in``b``|x,x',y)=\nonumber\\
\frac{1+(-1)^{a+a'}\cos(x+y)\cos(x'+y)}{4}
\eeqa
Hence, the joint probability distribution (\ref{jointProb}) is signalling from Bob to Alice! Is this a problem for Bohmian mechanics? Probably not, as the Bohmian particles's positions are assumed to be ``hidden''. Actually, it is already well-known that they have to be hidden in order to avoid signalling in Bohmian mechanics. Some may find this feature unpleasant, as it implies that Bohmian particles are postulated to exist ``only'' to immediately add that they are ultimately not fully accessible; but this is not new.

Consequently, defining a joint probability for the measurement outcomes $a$, $a'$ and $b$ in the natural way: 
\beqa\label{fullCorrel}
&P&(a,a',b|x,x',y)\equiv \nonumber\\
&P&(r_A\in``a``,r_A\in``a'``,r_B\in``b``|x,x',y)
\eeqa
can be done mathematically, but can't have a physical meaning, as $P(a,a',b|x,x',y)$ would be signaling. 

%Indeed, if Alice performs a position measurement after her first beam-splitter, then there is no longer a possibility for her to undo her measurements and perform a second one. And if she doesn't perform her first measurement, but only the second one, then she has no access to $r_A$, not even to which mode $r_A$ belongs to, hence she has no result $a$. Note that in both cases Alice and Bob get the no-signalling quantum correlation (\ref{correlab}) or (\ref{correla'b}), but they never have access to (\ref{fullCorrel}). 

\section{What is going on? Let's add a position measurement}\label{pointer}
%===========================================================
In summary, it is the identification (\ref{fullCorrel}) that confused the authors of \cite{CorreggiMorchio02,KiukasWerner09} and led them to wrongly conclude that Bohmian mechanics can't predict violations of Bell inequalities in experiments involving only position measurements. Note that  the identification (\ref{fullCorrel}) follows from the assumption {\bf H}, hence assumption {\bf H} is wrong. Every introduction to Bohmian mechanics should emphasize this. Indeed, assumption {\bf H} is very natural and appealing, but wrong and confusing.

%Consequently naive Bohmian mechanics is wrong and in non-naive Bohmian mechanicsthe identification (\ref{fullCorrel}) doesn't hold, i.e. a position measurement may provide a result that doesn't correspond to the particle's Bohmian position. 

To elaborate on this let's add an explicit position measurement after the first beam-splitter on Alice side. The fact is that both according to standard quantum theory and according to Bohmian mechanics, this position measurement perturbs the quantum state (hence the pilot wave) in such a way that the second measurement, labelled $x'$ on Fig. 4, no longer shares the correlation (\ref{PosCorrel}) with the first measurement, see \cite{Englert92, Vaidman03}.

\begin{figure}
\centering
\includegraphics[width=9cm]{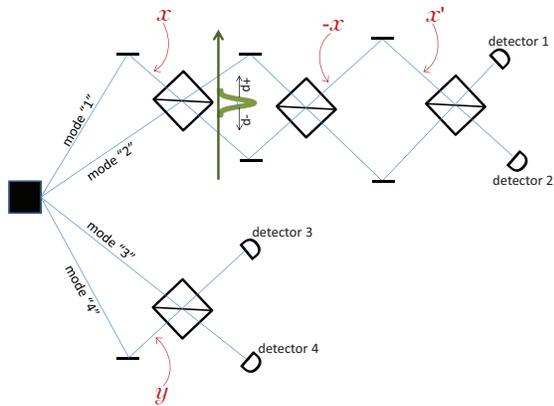}
\caption{We add a pointer that measures through which path Alice's particle propagates inbetween her first and second beam-splitter. The pointer moves up if Alice's particle went through the upper path, i.e. $r_A\in\,''1''$, and down if it went through the lower path, i.e. $r_A\in\,''2''$. Hence, by finding out the pointer's position one learns through which path Alice's particle went, i.e. one finds out Alice's first measurement result, though it all depends how fast the pointer moves. See text for explanation.}\label{fig4}
\end{figure}

Let's model Alice's first position measurement, labeled $x$ (i.e. corresponding to the input phase $x$), by an extra system, called here the pointer, initially at rest in a Gaussian state, see Fig. 4. One should think of the pointer as a large and massive system; note that it suffices to consider the state of the center of mass of the pointer. If Alice's particle passes through the upper part of the interferometer ($r_A\in\,''1''$), then the pointer gets a kick in the up direction and is left with a momentum $+p$; but if Alice's particle passes through the lower part of her interferometer ($r_A\in\,''2''$), then the pointer gets a kick $-p$. We chose p large enough that the two quantum states of the pointer $\ket{\pm p}$ are orthogonal, i.e. according to quantum theory we consider a strong (projective) which path measurement. Note however, that immediately after the pointer has interacted with Alice's particle the two Gaussians corresponding to $\ket{\pm p}$ overlap in space, hence no position measurement can distinguish them. It is only after some time that the two Gaussians separate in space and that position measurements can distinguish them. Since in Bohmian mechanics there are only position measurements, this implies that in Bohmian mechanics it takes some time for the pointer to measure Alice's particle.

Accordingly, if $p$ is large enough for the pointer to have moved by more than its spread by the time Alice's particle hits the second BS, then the pointer acts like a standard measurement, and the second position measurement $x'$ of Alice's particle is perturbed by measurement $x$, as discussed in the previous paragraph. But if $p$ is small enough, then, by the time the second measurement $x'$ takes place, the pointer almost didn't move. In this case, the second position measurement is not affected, \cite{Englert92, Vaidman03}, see also the appendix. However, it is now this second measurement $x'$ that perturbs the ``first'' one, i.e. perturbs measurement $x$. Indeed, because of the entanglement between Alice's particle and the pointer, if one waits long enough for the pointer to move by more than its spread and then reads the result of the ``first'' measurement out of this pointer, then one won't find the expected result: the second measurement perturbed the ``first'' one. I put ``first'' in quotes because in such a slow measurement the result is actually read out of the pointer after the second measurement took place.

This is very similar to the so-called surrealistic trajectories, see \cite{Englert92, Vaidman03}. In the appendix, I recall this counter-intuitive aspect of Bohmian mechanics.

%\section{Why Bohmian positions?}
%===============================
%But so, for what are the Bohmian positions good for? Indeed, one can't know their precise initial positions, as this would lead to signaling. And, as we have seen in this note, one can't know 2-time position correlations, and even less entire trajectories of these positions, as, again, it would lead to signaling. So, why should one introduce such hidden positions, as they have to be and remain hidden? Well, I guess it's a matter of taste. 

%Admittedly, Bohmian mechanics is a constructive  existence proof of nonlocal hidden variable models reproducing quantum predictions. Furthermore, it has some classical physics smell, though with a deep touch of nonlocality which changes almost everything: quantum entanglement still affects the pilot wave, hence hides the position as we have seen. 

%Many philosophers like Bohmain mechanics because one can postulate that the hidden Bohmian position is the real stuff - the ontology, sometimes even called the primitive ontology. Why not, provided one remembers that this primitive ontology can't be fully accessed, in strong contrast to positions of classical particles which, according to classical physics, can be revealed by merely looking at them.

\section{What about large systems?}
%==================================
So far so good. But let's now consider not single particles, but elephants. One of the nice features of Bohmian mechanics is that it makes no difference between microscopic and macroscopic systems: all systems are treated alike. The price to pay, as we illustrate below, is that all the strangeness of quantum physics at the microscopic level has to show up also at the macroscopic level. 

Let's consider two elephant in the state (\ref{psi_in}) corresponding to entangled elephants in modes 1\&4 superposed with elephant in modes 2\&3. Note that instead of elephant one may consider classical light pulses and replace in (\ref{psi_in}) the 1-photon state $\ket{1}$ by a coherent state $\ket{\alpha}$ with mean photon number $|\alpha|^2$ as large as desired: $\big(\ket{\alpha,0,0,\alpha}+\ket{0,\alpha,\alpha,0}\big)/\sqrt{2}$. The beam-splitters have to be replace by EBS - Elephant Beam Splitters - that split elephants: an incoming elephant emerges from a EBS in a superposition of elephant-transmitted and elephant-reflected. In the case of coherent states the transformation reads:
\beqa
\ket{\alpha,0}\rightarrow \big(\ket{\alpha,0}+i\ket{0,\alpha}\big)/\sqrt{2}\\
\ket{0,\alpha}\rightarrow \big(i\ket{\alpha,0}+\ket{0,\alpha}\big)/\sqrt{2}
\eeqa
Note that the above deeply differs from standard BS which correspond to $\ket{\alpha,0}\rightarrow\ket{\alpha/\sqrt{2},i\alpha/\sqrt{2}}$.

All the above single-particles story remains the same. Hence, in Bohmian mechanics, elephants' positions are also hidden, or at least not fully accessible. But this is puzzling as it means that when one ``looks slowly'' (as the pointer in  section \ref{pointer}, see also the appendix) at an elephant, one may see it where it is not. Indeed, according to Bohmian mechanics an elephant is where all the Bohmian positions of all the particles that make up the elephant are. But what does this mean if it doesn't correspond to where one sees the elephant? Bohmians may reply that one doesn't ``look slowly'' at elephants and that Elephant-Beam-Splitters don't exist. This is certainly true of today's technology, but there will soon be beam splitters for quantum systems large enough to be seen by the naked eye. And to avoid signaling it has to be impossible to ``see'' or find out in any way 2-time position correlations of such quantum systems, even when large.

Admittedly, Bohmian mechanics has the nice feature that it makes not difference between micro- and macro-worlds. But, accordingly and unavoidably, the quantum weirdness shows up at the macro-scale.

\section{Assumption {\bf H} revisited}
%=================================================================
Assumption {\bf H} is wrong. How should one reformulate it? Clearly, a position measurement doesn't {\it merely} reveal the Bohmian particle because:
\begin{enumerate}
\item a position measurement necessarily involves the coupling to a large system, some sort of pointer, and this coupling implies some perturbation. Hence the ``merely'' in assumption {\bf H} is wrong \cite{MaudlinPC15}.
\item depending how the coupling to a large system is done and on how that large system (the pointer) evolves a position measurement reveals some information about the Bohmian particle or not. Hence, not all measurements that, according to quantum theory, are position measurements, are also Bohmian-position measurements: some quantum-position measurements don't reveal where the Bohmian particle is.
\end{enumerate}

The first point above is very familiar to quantum physicists. However, it may take away some of the appeal of Bohmian mechanics. Indeed, the naive picture of particles with always well-defined positions is obsucred by the fact that these positions can't be ``seen'', actually one can not even ``merely see'' in which mode a Bohmian particle is. At the end of the day, Bohmian mechanics is not simpler, the promise of continuously well-defined position and the associated intuition is deceptive.

The second point listed above is interesting: one should distinguish {\it quantum-position} and {\it Bohmian-position} measurements. The later are measurements that provide information about the position of Bohmian particles. It would be interesting to find out how to characterize such Bohmian-position measurements without the need to fully compute all the Bohmian trajectories.

\section{Why Bohmian mechanics}
%==============================
From all we have seen so far, one should, first of all, recognize that Bohmian mechanics is deeply consistent and provides a nice and explicit existence proof of a deterministic nonlocal hidden variables model. Moreover, the ontology of Bohmian mechanics is pretty straightforward: the set of Bohmian positions is the real stuff. This is especially attractive to philosopher. Understandably so. But what about physicists mostly interested in research? What new physics did Bohmian mechanics teach us in the last 60 years? Here, I believe fair to answer: not enough! Understandably disappointing.

It is deeply disappointing that an alternative theory to quantum mechanics, a theory that John Bell thought should be taught in parallel to standard textbook quantum mechanics \cite{BellBohm}, didn't produce new physics, not even inspirations for new ideas to be tested in the lab (though see \cite{Valentini}). How could that be? Some may conclude that not enough people worked on Bohmian mechanics. But tens or hundreds of passionate researchers work on it during decades. Some may conclude that this lack of new ideas proves that Bohmian mechanics is a dead end. But how could a consistent theory, empirically equivalent to quantum theory, have no future?

Let me suggest some possible answers to the above puzzle, though only partial answers. I am afraid that almost all the research in Bohmian mechanics over the last decades remained trapped in an exceedingly narrow view and worked only on highly specific problems of interest only to their Bohmian community. I believe this is especially disappointing as there were several interesting open problems that Bohmian-inspired ideas could have addressed. The positive side is that it is likely that there still are interesting open problems waiting for open mind researchers.

Let me illustrate some of the ideas I believe Bohmian mechanics should have triggered. This list is obviously subjective, important is only that it is not empty. Bohmian mechanics, like quantum theory, is in deep tension with relativity theory. I know of Bohmians who claim that it is obvious that any non-local theory, Bohmian or not, requires a privileged universal reference frame. I also know of Bohmian who claim that it is obvious that Bohmian mechanics can be generalized to a relativistic theory (though, admittedly, I never understood their model). But I know of no Bohmians who tries to get inspiration from their theory and its tension with relativity to try to go beyond Bohmian mechanics, as illustrated in the next two paragraphs. 

According to Bohmian mechanics, particles ``make decisions'' at beam-splitters in the sense that after a beam-splitter the particle is definitively in one of the output modes. Admittedly, this is not a real decision as everything is determined by the initial state of the particle and of all other systems entangled with the particle. But let me keep this inspiring terminology. Accordingly, and following Suarez, we call such beam-splitters {\it choice-devices} \cite{SuarezChoicedevice}. Such choice-devices take into account everything in their past. Now, a natural assumption inspired by the sketched description, is that the past is not merely the past light cone, but all the past in the inertial reference frame of the choice device. This idea led Suarze and Scarani to suggest that one should test situations in which several choice-devices, e.g. several beam-splitters, are in relative motion such that what is the past for one choice-device may differ from the past of another choice-device \cite{SuarezScarani}. This has the very nice feature (at least for researchers in physics) that it leads to experimental predictions that differ from standard quantum predictions and that can be experimentally tested. Hence, this brings Bohmian-inspired ideas to physics. This has been tested in my lab, the result proves the idea wrong \cite{SZGS02}. Appealing but wrong. Wrong but new and testable.

Another Bohmian-inspired idea follows directly from an observation by Hiley and Bohm \cite{BohmHiley}: ``it is quite possible that quantum nonlocal connections might be propagated, not at infinite speeds (as in standard Bohmian mechanics), but at speeds very much greater than that of light. In this case, we could expect observable deviations from the predictions of current quantum theory (e.g. by means of a kind of extension of the Aspect-type experiment)". Again this can be experimentally tested \cite{pla00, Salart, Cocciaro10, China}. The results put lower bounds on this hypothetical faster-than-light-but-finite speed influence, something like 10'000 to 100'000 times the speed of light. Aspect-type experiments between two sites can only, either find that hypothetical speed or set lower bounds on it. But recently we could demonstrate that by going to more parties one can prove that either there is no such finite-but-superluminal speed or that one can use it for faster than light communication using only classical inputs and output (i.e. measurement settings and results) \cite{Bancal12,Tomer13}.

I am confident that Bohmian mechanics and other alternative views on quantum mechanics will inspire further ideas that will lead to nice experiments bounding possibly extensions of quantum theory. The real question is whether the Bohmian community will contribute to them.

\section{Conclusion}
%===================
In conclusion, naive Bohmian mechanics - that include assumption {\bf H} - is wrong. Yet, Bohmian mechanics is deeply consistent. Position measurements perturb the system, even in Bohmian mechanics. Hence, the existence of two-time position correlations is not in contradiction with possibly violations of Bell inequalities.

Generally, position measurements reveal some information about Bohmian positions, but never full information and sometimes none at all. A simply and handy criteria as to when a measurement highly correlates Bohmian positions of a particle under test to the position of the center of mass of some large pointer is still missing.

Bohmian mechanics is attractive to philosophers because it provides a clear ontology. However, it is not as attractive to researchers in physics. This is unfortunate  because it could inspire courageous ideas to test quantum physics.

\appendix\section{Appendix: Slow position-measurements in Bohmian Mechanics}
%=========================================================
{\bf Bohmian trajectories in a semi-interferometer}\\
%======================================================
Let us consider a ``half Mach-Zehnder'' interferometer, that is a Mach-Zehnder interferometer in which the second beam-splitter is removed, see Fig. 5. We call such a circuit a semi-interferometer. After the beam-splitter any particle entering from the input mode is in a superposition of modes 1 and 2: $\ket{1}\rightarrow\big(\ket{1,0}+\ket{0,1}\big)/\sqrt{2}$, with possibly a relative phase irrelevant in semi-interferometers.

\begin{figure}
\centering
\includegraphics[width=9cm]{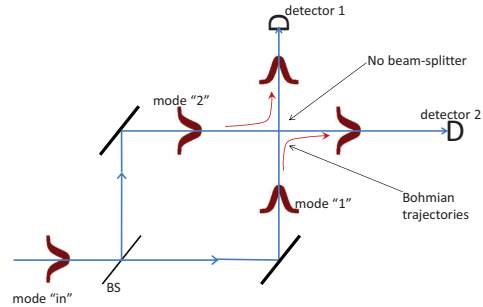}
\caption{A Bohmian particle and its pilot wave arrive on a Beam-Splitter (BS) from the left in mode ``in''. the pilot wave emerges both in modes 1 and 2, as the quantum state in standard quantum theory. Modes 1 and 2 meat again, but there is no beam-splitter at this meeting point. Nevertheless, the Bohmian trajectories bounce at this point as indicated by the red arrows. Intuitively this can be understand because the evolution equation of the Bohmian position is a first order differential equation, hence Bohmian trajectories never cross each other. This intuition is confirmed by numerical simulations.}\label{fig5}
\end{figure}

According to quantum theory if detector 1 clicks, then the particle went through mode 1. This should be interpreted as ``if one adds position measurements in modes 1 and 2, then there is a 100\% correlation between detector 1 and the position measurement in mode 1 (and similarly for detector 2 and the position measurement in mode 2)''.

According to Bohmian mechanics things are different. If detector 1 clicks, then the particle went through mode 2, in sharp contrast to the quantum retro-diction. However, the interpretation is also totally different. According to Bohmian mechanics, particles follow continuous trajectories and the interpretation here is that, if detector 1 clicks, then the Bohmian particle followed mode 2.

In order to reconcile both views, let's add position measurements in modes 1 and 2.

{\bf Position Measurements in modes 1 and 2}\\
%===============================================
In order to describe position measurements in both modes we add two pointers, each initially at rest, denoted $\ket{p_j=0}$, that we locally couple to modes 1 and 2 in such a way that if the particle is in mode $j$ then the correspond pointer gets a momentum kick $k$, resulting in state $\ket{p_j=k}$, while the other pointer is left unaffected, see Fig. 6. The joint particle-pointers state after the two local interactions thus read: $\big(\ket{1,0}\ket{p_1=k}\ket{p_2=0}+\ket{0,1}\ket{p_1=0}\ket{p_2=k}\big)/\sqrt{2}$. Note that the pointers are in some localized (e.g. Gaussian) states, the kets only indicate the mean momenta.

Let us emphasize that the pointer should be thought of as large, massive and consisting of many internal degrees of freedom; in short it is a ``macroscopic'' object and the result of the position measurement can merely be read of the pointer's position: if the pointer moved, then it indicates that it has detected the presence of the particle, if the pointer hasn't moved, then this indicates that the particle went the other way. Such a formalization of position measurements applies both to quantum and Bohmian theories.

Note that in order for the pointer to indicate an unambiguous result one has to wait long enough for the pointer to have moved by much more than it's spread $\Delta x$ and the kick has to be large enough, $k>>\hbar/\Delta x$.

According to quantum theory, if detector 1 clicks, then pointer 1 got a kick and thus moves, while pointer 2 rests in state $\ket{p=0}$. However, the situation as described by Bohmian mehcanics is more interesting.

First, consider the case that the kick $k$ is so large that the pointer, if kicked, move by more than it's spread before the two modes 1 and 2 cross at the place of the ``missing beam-splitter''. In this case, the particle and the kicked pointer get entangled and this modifies the Bohmian trajectory of the particle. According to this modified trajectory and in full agreement with quantum predictions, if detector 1 clicks, then Bohmian mechanics predicts that it is pointer 1 that moved (including the Bohmian position of pointer 1). Note that in this situation Bohmian trajectories can apparently cross each other, because the trajectory actually happens in a higher dimensional space and it is only its shadow in our space that cross.

Next, consider the case that the kick $k$ is not that large and that, by the time modes 1 and 2 cross, the pointer barely moved. Then Bohmian mechanics predicts that if detector 1 clicks, the particle went through mode 2 (as if there were no pointer); however, if one waits long enough for the pointer to eventually move by more than it's spread, then one finds that it is pointer 1 that moves. Accordingly, in the case of ``slow pointers'' the pointer indicates where the Bohmian particle was not. This is surprising, at least to physicists used to quantum theory. But this is how Bohmian mechanics describes the situation and one should add that there is nothing wrong with this description in the sense that all observable predictions are in agreement with quantum predictions.

Finally, we investigated numerically intermediate cases in which, at the time modes 1 and 2 cross, the pointer moved but not much. We find that in such cases, some trajectories of the particle bounce in the region of the mode crossing, while other trajectories go through the crossing region more or less in straight lines. Accordingly, conditioned on detecor 1 clicking, there is a chance that the particle went through mode 1 and a complementary chance that it went through mode 2, depending on the precise value of the kick $k$ and the exact initial position of the Bohmian particle.

\begin{figure}
\centering
\includegraphics[width=11cm]{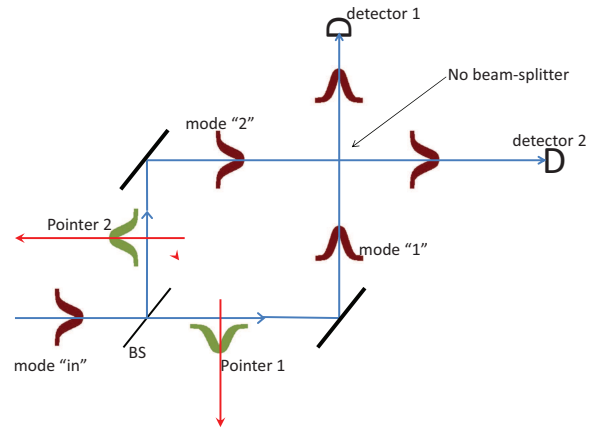}
\caption{Semi-interferometer with two macroscopic pointers locally coupled to modes 1 and 2. The pointers are initially at rest, $\ket{p_j=0}$, but when detecting a particle they get a kick and end in a quantum state with momentum $k$: $\ket{p_j=k}$.}\label{fig6}
\end{figure}

Note that it suffices that one of the two detectors moves fast to avoid that the Bohmian trajectories bounce in the crossing region. Actually, it suffices that there is only one detector; then, dependent on the kick received by this single detector, the Bohmian trajectories bounce or not.

{\bf Conclusion}\\
%===================
What is a position measurement? Quantum theory has a clear answer to this question. However, in Bohmian mechanics there are two possible definitions. First, the natural one: a {\it Bohmian-position} measurement is any interaction between the particle under test and a macroscopic device (e.g. a pointer) that fully correlates the Bohmian position of the particle immediately before the interaction took place to the final state of the device (e.g. final position of the pointer). Next a quantum inspired one: Anything that is a position measurement according to quantum theory (i.e. represented by the position operator $q$ or a function of it) is also a {\it quantum-position} measurement in Bohmian mechanics. 

Hence, in Bohmian mechanics one should distinguish Bohmian-position measurements and quantum-position measurements. In most situations both types of position measurements coincide. But there are cases, as the slow pointer described in this note, where the two concepts differ deeply.

When one says that Bohmian mechanics makes the same predictions as quantum theory as long as all measurements, at the end of the day, reduce to position measurements, one refers to quantum-position measurements. A said this may differ from Bohmian-position measurements, hence the Bohmian trajectories may differ from quantum expectations. This is surprising to quantum physicsits, but one should emphasize that there is nothing  wrong with that: different theories lead to different pictures of reality.

\small{
\section*{Acknowledgment} A preliminary version of this note was presented at a workshop at the ETh- Zurich in Octobre 2014 organised by Gilles Brassard and Reneto Renner. There, participants drew my attention to reference \cite{Englert92}. The present version profited from comments by Sandu Popescu who drew my attention to \cite{Vaidman03} and by Michael Esfeld and Tim Maudlin. Work partially supported by the COST Action {\it Fundamental Problems in Quantum Physics} and my ERC-AG MEC.
}

\end{document}